\documentstyle[emulateapj,pstricks]{article}   

\def\etal{{\it et al.\ }}
\def\ie{{\it i.e.\ }}

\def\Msun{~M_{\odot}\ }

\def\Mhsun{h^{-1}{\ }{\rm M_{\odot}}\ }

\def\Mpch{h^{-1}{\rm Mpc}}
\def\hMpc{h^{-1}{\rm Mpc}}

\def\hkpc{h^{-1}{\rm kpc}}

\def\etal{ et al.\ }

\def\M10{{\times 10^{10} M_{\odot}\ }}

\def\simov#1#2{\lower .5pt\vbox{\baselineskip0pt \lineskip-.5pt
       \ialign{$\mathnew#1\hfil##\hfil$\crcr#2\crcr\sim\crcr}}}

\def\today{\ifcase\month\or
  January\or February\or March\or April\or May\or June\or
  July\or August\or September\or October\or November\or
  December\fi
  \space\number\day, \number\year}

\begin{document}
\slugcomment{{\em submitted to the Astrophys.J.}}
\lefthead{MERGING HISTORY}
\righthead{GOTTL\"OBER, KLYPIN \& KRAVTSOV}

\title
{Merging history as a function of halo environment}

\author{Stefan Gottl\"ober}
\affil{Astrophysikalisches Institut Potsdam,
An der Sternwarte 16, D-14482 Potsdam, Germany}
\author{Anatoly Klypin} 
\affil{Department of Astronomy, New Mexico State University,
 Las Cruces,  NM 88001, USA\\}
\author{Andrey V. Kravtsov\footnotemark\\}
\affil{Department of Astronomy, The Ohio State University, 
140 West 18th Ave.,\\ Columbus, OH 43210-1173, USA\\}

\begin{abstract}

  According to the hierarchical scenario, galaxies form via merging
  and accretion of small objects. Using $N$-body simulations, we study
  the frequency of merging events in the history of the halos.  We
  find that at $z \la 2 $ the merging rate of the overall halo
  population can be described by a simple power law $(1+z)^{3}$.  The
  main emphasis of the paper is on the effects of environment of halos
  at the present epoch ($z=0$). We find that the halos located inside
  clusters have formed earlier ($\Delta z \approx 1$) than isolated
  halos of the same mass. At low redshifts ($z<1$), the merger rate of
  cluster halos is 3 times lower than that of isolated halos and 2
  times lower than merger rate of halos that end up in groups by
  $z=0$.  At higher redshifts ($z\sim 1-4$), progenitors of cluster
  and group halos have 3--5 times higher merger rates than isolated
  halos. We briefly discuss implications of our results for galaxy
  evolution in different environments.

\end{abstract}

\section{INTRODUCTION}
\label{intro}

\footnotetext{Hubble Fellow}
A significant fraction of mass in the universe
is believed to be in the form of dark matter (DM).  According to the
standard theoretical paradigm of structure formation, small-mass DM
perturbations collapse first and the resulting objects then merge to
form increasingly larger DM halos.  Baryonic matter (gas) is assumed
to follow the gravitationally dominant dark matter.  Galaxies, thus,
could have been formed within dense DM halos when the infalling gas
reaches sufficiently high overdensities to cool, condense, and form
stars.  The most convincing observational evidence for substantial
amounts of dark matter even in the very inner regions of galaxies
comes from HI studies of dwarf and low surface brightness galaxies.
The gravitational domination of DM on the scale of galaxy virial
radius implies that collisionless simulations can be used to study the
formation of the DM component of galaxies.

Interactions  between halos, such   as mergers, collisions, and  tidal
stripping,  are thought to   play a crucial role  in  the evolution of
galaxies.   In   particular, there  is    a substantial evidence  that
elliptical galaxies may have  formed   by mergers of disk  systems  (e.g.,
Barnes   1999). Observations of  faint   distant systems indicate that
interaction rate rapidly increases with redshift (e.g., Abraham 1999).
Intuitively, one could expect that the merging rate of galaxies should
depend  on environment (in    particular,  on the  local  density  and
velocity dispersion).  For example, Makino  \&  Hut (1997) found under
some simplifying  assumptions that  the  merging rate in  clusters  is
proportional  to $n^2\sigma^{-3}$ where $n$   is the number density of
galaxies in the cluster and $\sigma$ is their one dimensional velocity
dispersion of galactic velocities.  Since the environment changes with
time one could also  expect dramatic changes  in the evolution of  the
merging rate.

In order to study the evolution of the merging rate and its dependence
on   environment one must   follow   the evolution    of  halos in   a
representative cosmological   volume.  Moreover,  the simulation  must
have  sufficiently   high  mass and   force  resolutions. Insufficient
resolutions leads to structureless virialized halos instead of systems
similar  to observed groups  and  clusters of  galaxies with wealth of
substructure. This effect is well   known as the overmerging   problem
(e.g., Moore \etal 1996, Frenk \etal 1996, Klypin \etal 1999).

Cosmological scenarios with cold dark matter (CDM) alone cannot explain
the structure formation both on small and very large scales.  Variants
of the CDM model with a non-zero cosmological constant, $\Lambda$, have
proven to be very successful in describing most of the observational
data at both low and high redshifts. Moreover, from a recent analysis
of 42 high-redshift supernovae Perlmutter \etal (1999) found direct
evidence for $\Omega_{\Lambda}= 0.72$, if a flat cosmology is assumed.
Also, from the recent BOOMERANG data Melchiorri \etal (1999) found
strong evidence against an open universe with $\Lambda = 0$. For our
study we have chosen a spatially flat cosmological model with a
cosmological constant $\Omega_{\Lambda}= 0.7$ and the present-day
Hubble constant of $H_0=70$~km/s/Mpc.

The goal of this study is to determine  the distribution with redshift
of merging  events for  halos   which exist  at $z=0$. We   study this
merging rate and its dependence on environment of  halos at $z=0$. The
paper is  organized  as follows. In the   next  section we  define the
merging events studied in this paper. In \S~3 we describe the
cosmological model and  the numerical simulation.  We briefly describe
our  halo  finding algorithm,  the  definition of  environment and the
detection of progenitors  of   halos. The technical details   of these
procedures are   presented   in the  Appendix.  We  use   the extended
Press-Schechter formalism to  test our  procedure. In \S~4
we discuss  the merging of halos  found in  the simulation and compare
our results  with  observations. In   \S~5 we summarize  our
results and briefly discuss their implications.

\section{MERGING OF HALOS}\label{merg}

According to the hierarchical scenario, galaxies and the dark matter
halos associated with them have been formed in a process of merging
with other halos and accretion of small objects. Here merging denotes
the coalescence of two objects with comparable masses whereas
accretion means the infall of objects with masses much smaller than the
mass of the accreting object. Obviously, there is no sharp distinction
between the two processes.

During the formation of every halo there are events (let us name them
{\em major mergers}), when mass of the halo increases substantially
over a short period of time. Such events are very important because
they can lead to dramatic changes in the structure of dark matter
halos and galaxies they harbor. For example, the increase in mass
leads to the change of potential and, likely, density structure of the
dark matter halo. One may expect even more dramatic changes for the
baryonic component. Infalling objects may, for example, damage or even
destroy stellar disk. The inflow of material may also serve as a
source of fresh gas and may therefore induce increase in star
formation rate.  At the same time, collisions between halos may result
in shock heating of the gas, which would tend to delay or prevent star
formation for some period of time.

Impact of a major merger on the structure of a halo or a galaxy likely
depends on a particular configuration of the merging event: one
massive infalling object or accretion of many small-mass objects,
gas-rich or gas-poor mergers, etc. Nevertheless, it is reasonable to
expect that accretion of many small halos is as damaging as merging
with one massive satellite of the same total mass. Thus, it is logical
to define major merger event as an accretion event in which mass of a
halo increases substantionally (say, by more than 20\% -- 30\%) over a
short period of time (e.g., one dynamical time of the halo), as opposed
to a single merger with a massive halo.

There is another issue related to the major merger definition and
statistics.  One can consider mergers in a population of {\em all}
halos present at a given redshift. In this case, one counts all
merging events and divides the count by the total number of halos.
This gives an estimate of the merging rate. The procedure should be
used, for example, if one compares the frequency of close pairs with
theoretical predictions. In this paper we consider a different
statistics: {\em we study merging history of present day halos}.
Namely, we ask the following question: what is the probability that
the most massive progenitor of a $z=0$ halo identified at redshift $z$
had a major merger at this redshift?  At relatively small redshifts
($z < 1$), the merging rate of galaxy-size halos is low and the
differences between the merger rates defined in these two different
ways are small.  At higher redshifts the differences may become
substantial.

We are also interested in the effect of the environment of halos on
their merging rate. The environment of a given halo changes with time.
An isolated halo may fall into a group or cluster and groups, in their
turn, may then be accreted onto clusters. In this paper, we study the
differences between merging histories of halos residing in different
environments at $z=0$.

The ultimate outcome of merging or accretion event 
depends on both the time interval and on the fractional mass increase 
within this time interval. Observationally, the
time scale of merging is the time interval for which traces of the
event can be observed.  Physically, this time scale is of order of
the dynamical time of accreting halo. A reasonable lower limit on the merging 
time scale is the crossing time of the halo defined as the ratio of the 
halo radius, $R$, to the typical accretion velocity, $V$:
\begin{equation}
 t_{\rm cross}\approx 1{\ \rm Gyr}\left(\frac{R}{200{\ }{\rm kpc}}\right)
\left(\frac{V}{200{\ }{\rm km\ s^{-1}}}\right)^{-1}.
\end{equation}
The crossing time is approximately equal to one gigayear for a wide
range of halo masses. We conclude therefore that it is reasonable to
consider time interval as large as 500 Myrs in analyzing the
simulations.  For our analysis we have used the simulation outputs 
at 25 time moments. Although the time intervals between two stored
moments differ slightly, the mean interval is about 0.5 Gyr. 

In \S~3.1 we will use the extended Press-Schechter formalism to
test the effects of the different choices for the time interval. We
find that our results do not change significantly if we vary the time
interval from 0.1 to 0.5~Gyr. We have chosen a minimum fractional mass
increase of 25\% to define a major merger. The resulting merging rate
depends slightly on the choice of this value as discussed in
\S~4.  To summarize, in the remainder of the paper the
major mergers are defined as accretion events in which mass $M_1$ of
the most massive progenitor of a present-day halo increases by more
than 25\%: $(M_2 - M_1)/M_2 > 0.25$, where $M_2$ is halo mass after
merging.

In general, during the evolution of a halo in the simulation its mass
increases due to accretion and merging. However, interacting halos may
also exchange and lose mass. In particular, tidal stripping (and thus
mass loss) becomes important in the dense environment of clusters
(Klypin \etal 1999; Gottl\"ober \etal 1999a). Depending on the
environment of the halo at $z=0$, we have divided our sample into three
subsamples: isolated halos, halos in groups, and halos in clusters.

Merging rates estimated using the extended Press-Schechter (EPS; Bond
et al. 1991; Lacey \& Cole 1993) formalism are expected to be close to
those in numerical simulations. We use the EPS formalism to test
robustness of our assumptions and parameters used in the analysis
(e.g., the time interval of the merging).  It should be noted that
halos are treated differently in simulations and in the EPS formalism.
By definition, in the EPS formalism the halos are isolated and their
mass can only increase with time.  Substructure (subhalos inside a
larger halo) is not considered by the EPS. Let us consider a merger of
two isolated halos with substructure.  From the point of view of the
EPS formalism this is a single merging event: the mass of the
resulting merger product is considerably higher than the mass of the
individual merging systems. In the simulation we would detect the mass
growth for the most massive progenitor of the merged halo, but not for
the individual subhalos present as substructure. In fact, the subhalos
may even lose some mass due to the tidal stripping. To summarize, an
accretion event that should be classified as a merger on mass-scale of
a group or cluster, may not have a corresponding merger on the
mass-scale of the galaxy-size halos belonging to the group.  This
example shows that somewhat different results must be expected for the
merging rates of halos in the EPS formalism and in the simulation.
Nevertheless, the general behavior is expected to be the same and the
EPS formalism is a powerful method to test the assumptions made when
studying the merging rate of halos in a numerical simulation.

\section{COSMOLOGICAL MODEL AND NUMERICAL SIMULATION}
\label{simu}

We study a flat CDM cosmological model with a non-zero 
cosmological constant ($\Lambda$CDM). The model has following
parameters: $\Omega_0=1-\Omega_{\Lambda}=0.3$; $\sigma_8=1.0$; $H_0=70$
km/s/Mpc. It is normalized in accord with the four year {\sl COBE} DMR
observations (Bunn \& White 1997) and observed abundance of galaxy
clusters (Viana \& Liddle 1996).  The age of the universe in this model
is $\approx 13.5$ Gyrs.

The main goal of this study is the evolution of both isolated halos and
halos located inside virial radii of larger group- and cluster-size
systems.  This requires high force and mass resolution of the
simulation. The force and mass resolution required for a simulated halo
to survive in the high-density environments typical of groups and
clusters is $\sim 1-3$ kpc and $\sim 10^9{\rm M_{\odot}}$, respectively
(Klypin \etal 1999).  We use the Adaptive Refinement Tree (ART)
$N$-body code (Kravtsov, Klypin \& Khokhlov 1997) to follow the
evolution of $256^3$ dark matter particles with the range in spatial
resolution of $32,000$. With the required resolution we can simulate
the formation of halos in a box of $60 \hMpc$.  With $256^3$ dark
matter particles the particle mass is $1.1 \times 10^9 h^{-1} {\rm
M_{\odot}}$. We reach the force resolution of $2 \hkpc$. In the box of
this size there are sufficiently large number of halos in different
environments. This allows us to study the merging rate of halos in
different environments.

\begin{figure*}[ht]
\pspicture(0,0)(18.5,19.5)

\rput[tl]{0}(2.5,21.){\epsfxsize=13cm
\epsffile{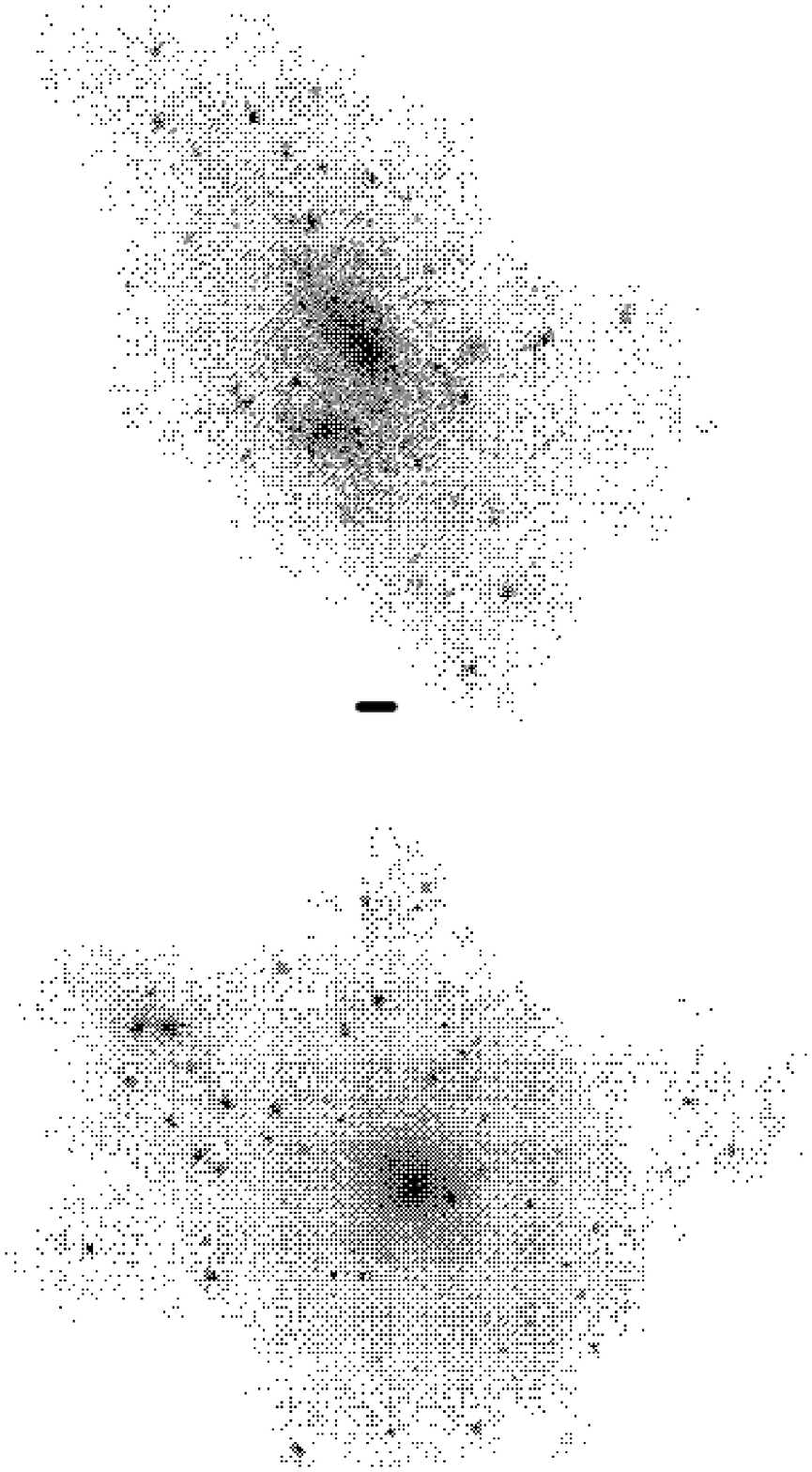}}

\rput[tl]{0}(0.,1.5){
\begin{minipage}{18.5cm}
  \small\parindent=3.5mm {\sc Fig.}~1.--- Cluster-size halos at two
  redshifts. {\em Top:} two group-size halos with a total mass of $9.2
  \times 10^{13} \Mhsun $ are merging at $z=1$.  Two largest halos
  close to the center will merge and produce one halo at the center of
  the final cluster. {\em Bottom:} the product of the merger shown in
  the top panel (mass $1.5\times 10^{14} \Mhsun$) at $z=0$.  The bar
  between the panels corresponds to a length of $100 \hkpc$
  (comoving); the comoving extent of the shown particle distribution
  is about $3\Mpch$.
\end{minipage}
}
\endpspicture
\end{figure*}

The evolution of an isolated halo (galaxy-, group-, or cluster-size),
whose mass grows due to accretion and merging, is relatively simple.
This mass growth is well described by the extended Press-Schechter
model. However, in this analysis we are interested not only in the
evolution of isolated halos but also in the evolution of {\em subhalos}
located within groups or clusters, \ie in the evolution of {\it
substructures} of bigger isolated objects. In fact, isolated halos and
subhalos in groups or clusters evolve differently (cf.  Gottl\"ober
\etal 1999a). For example, the latter may loose mass due to tidal
interaction when falling into the group or cluster or merge with the
host halo if their orbit decays due to dynamical friction.

The bottom panel of Fig.~1 shows a typical medium-size
cluster of mass $1.5 \times 10^{14} \Mhsun$ and diameter of about $ 3
\hMpc$ (extent of the shown particle distribution) at $z=0$. The
figure shows that the final halo contains many subhalos. The figure
shows all DM particles of the cluster which are linked using the
friends-of-friends (FOF) algorithm with the linking length of 0.2
times the mean interparticle separation.  (This linking length
approximately corresponds to the virial overdensity).  The particles
are colored on a gray scale according to the logarithm of the local
density at particle position smoothed over a sphere of the comoving
radius of $10 \hkpc$.  This cluster has formed through a merger of two
massive groups; at $z=1$ (top of figure), the merger is still in
progress. The two largest halos apparent at $z=1$ merged into one
central object by $z=0$.  The galaxy size halos that can be seen
within this cluster were formed well before the cluster formation in a
region of high density.  At $z<1$, the cluster grows further through
relatively mild accretion of dark matter and DM halos: mass increased
by a factor 1.6 from $z=1$ to $z=0$. With the high resolution of the
simulation we can follow the evolution of each halo (with mass above a
certain threshold determined by the mass resolution of the simulation)
from the moment of its formation until $z=0$.

\begin{figure*}[ht]
\pspicture(0,0)(18.5,19.5)

\rput[tl]{0}(2.5,21.5){\epsfxsize=13.5cm
\epsffile{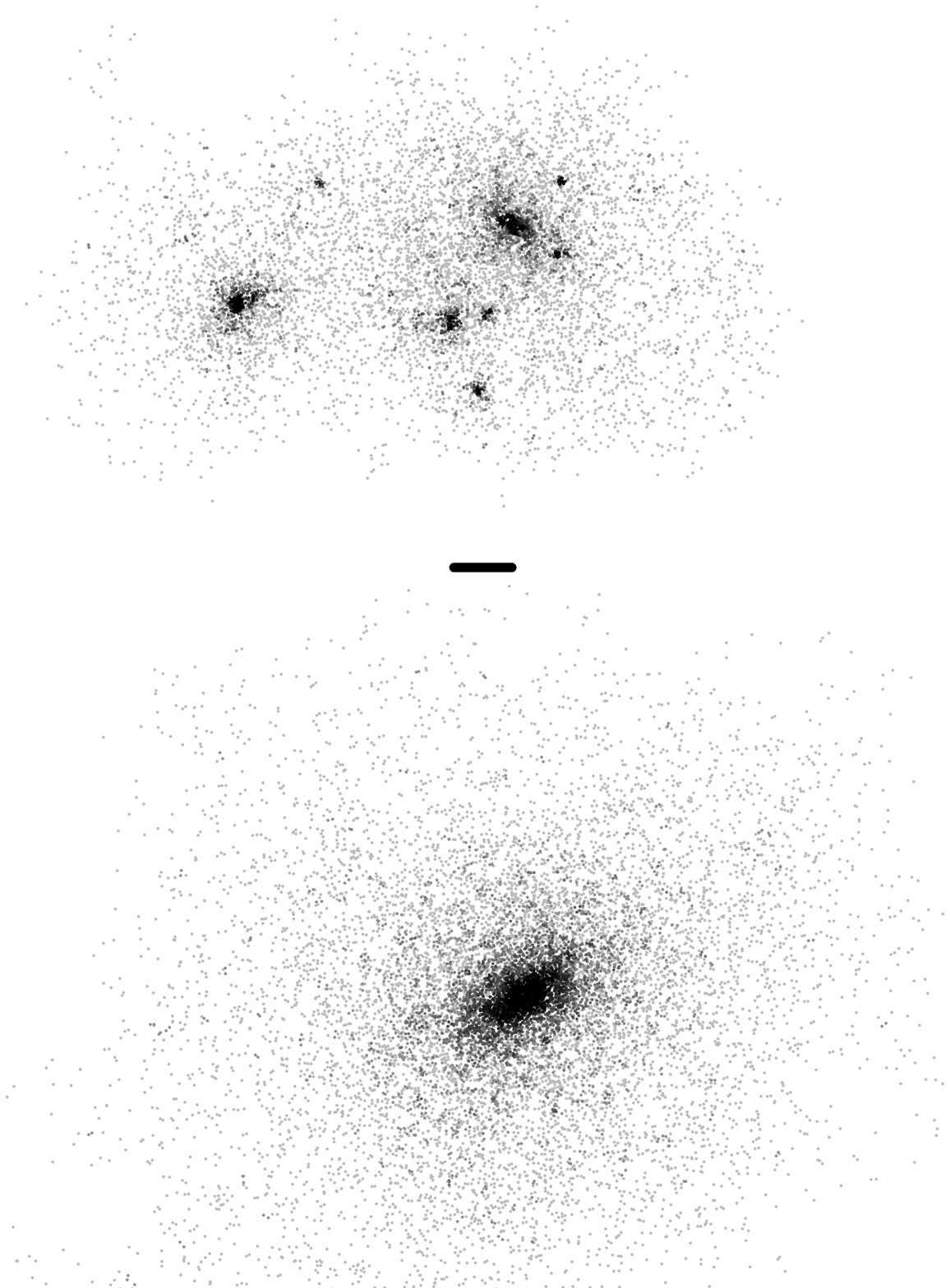}}

\rput[tl]{0}(0.,1.5){
\begin{minipage}{18.5cm}
  \small\parindent=3.5mm {\sc Fig.}~2.--- An example of (rare) merging
  of a group of halos to an isolated galaxy-size halo. {\em Top:} a
  group of halos of total mass $9.4 \times 10^{12} \Mhsun $ at $z=1$.
  {\em Bottom:} at $z=0$ this group has merged to form an isolated
  halo of $1.5 \times 10^{13} \Mhsun$.  No subhalos with maximum
  circular velocities $>100$ km/s has survived. The bar between the
  panels corresponds to a length of $100 \hkpc$ (comoving).
\end{minipage}
}
\endpspicture
\end{figure*}

Fig.~2 shows the extreme case  when a large halo does not
have substantial substructure  within  virial radius.   In the  bottom
panel of  Fig.~2 we  show an  isolated  halo of mass $1.5
\times 10^{13}  \Mhsun$. The extended  dark matter halo has a diameter
of about $ 1 \hMpc$. At redshift $z=1$  the progenitor of this halo is
a group of small-mass halos with the total mass of $9.4 \times 10^{12}
\Mhsun $ and a size of about $1.3 \hMpc$. This is an interesting but a
rare case: in  the simulation  we found  20 halos  ($>10^{12} \Mhsun$)
that were classified as isolated at  $z=0$ (no subhalos of $v_{circ} >
100$ km/s), but  whose $z=1$ progenitors were  groups of four to seven
members with   $v_{circ} > 100$  km/s.   Observed counterparts of such
merged halos could be the massive isolated ellipticals with group-like
X-ray halos (see Mulchaey \& Zabludoff 1999; Vikhlinin \etal 1999).

\subsection{Finding halos at different redshifts}
\label{def_samp}
Identification of halos in dense environments and reconstruction of
their evolution is a challenge. The most widely used halo-finding
algorithms, the friends-of-friends (FOF) and the spherical overdensity,
both discard ``halos inside halos'', \ie, satellite halos located
within the virial radius of larger halos. In order to cure this, we
have developed and used two algorithms to find halos: the {\em
hierarchical friends-of-friends} (HFOF) and the {\em bound density
maxima} (BDM) algorithms (Klypin \etal 1999). The HFOF algorithm uses
a set of different linking lengths in order to identify the
substructures of large DM halos as ``halos inside halos''. The BDM
algorithm does the same by identification of all local density maxima
and following the density profiles starting at these points. 

The HFOF and BDM algorithms are complementary. Both of them find
essentially the same halos above a reasonable mass threshold ($\ga 30$
particles). Therefore, we believe that each of them is a stable
algorithm which finds in a given dark matter distribution the DM
halos.  The advantage of the HFOF algorithm is that it can handle
halos of arbitrary, not only spherically symmetric, shape. The
advantage of the BDM algorithm is that it describes better the
physical properties of the halos because it separates background
unbound particles from the particles gravitationally bound to the halo
and constructs density and velocity profiles for each halo.

It is difficult and usually ambiguous to find mass of a halo located
within a larger halo. The formal virial radius of such a halo is equal
to the bound system's virial radius. If necessary, we define halo mass
as mass within its tidal or truncation radius defined as a radius
where the halo density profile starts to flatten. We try to avoid the
problem of mass determination by assigning not only the mass to a
halo, but finding also its maximum circular velocity
$v_{circ}=\sqrt{GM(<R)/R}|_{max}$. Numerically, $v_{circ}$ can be
measured more easily and more accurately then the mass. Beside of
being more stable numerically, the circular velocity is also more
meaningful observationally.

For  our  analysis we  need a  halo  sample which   is as  complete as
possible but does not contain any fake  halos. Recently, we have shown
that the halo samples constructed from the simulation used here do not
depend on  the numerical parameters of  the halo finder for halos with
$v_{circ}  {_ >\atop{^\sim}} 100$  km/s  (Gottl\"ober \etal 1999a). At
$z=0$,  we decided to  be even more restrictive  and limit analyses to
halos with a circular velocity of $v_{circ}  > 120$ km/s which contain
more than 100 bound particles within a radius of $100 \hkpc$. To avoid
misidentifications we have required that each halo of our sample has a
unique progenitor  at  the last   five time  steps (see  Appendix  for
details).  At $z=0$  our halo sample consists  of 4193 halos. The halo
number density, $0.019 h^3 {\rm Mpc}^{-3}$, roughly corresponds to the
number density of galaxies with $M {_ <\atop{^\sim}} -18.5$ in the Las
Campanas redshift survey (Lin \etal 1996).

\newpage
\subsection{Definition of environment}
\label{def_env}

As mentioned above, the goal  of this paper  is  to study the  merging
history  of the halos  as a function  of their  environment at present
($z=0$). To characterize the environment  we have run friend-of-friend
analysis of the simulation outputs with a linking  length of 0.2 times
the mean interparticle distance.  The FOF algorithms, thus, identifies
clusters  of DM particles with  average overdensity of  about 200. The
virial  overdensity in the  $\Lambda$CDM model  under consideration is
about 330  which  corresponds to  a  linking length   of about 0.17.   
Therefore, the  objects which we  find have  a slightly  larger extent
than the objects at virial overdensity. We  have increased the linking
length to account (at  least partially) for the  halos gravitationally
bound to the  cluster halo but located just  outside  at the epoch  of
identification.

For each of the identified halos, we find a host halo if such host
exists (see Appendix). We call the halo  {\em isolated}, if it does not
belong to any higher-mass host. We call  the halo a {\em cluster}
if it  belongs to  a particle cluster  with  a total mass  larger than
$10^{14} h^{-1} {\rm M_{\odot}}$.  Finally,  we identify {\em   group}
{\pspicture(0,0)(11.5,12.5)

\rput[tl]{0}(0,12.5){\epsfxsize=8.5cm
\epsffile{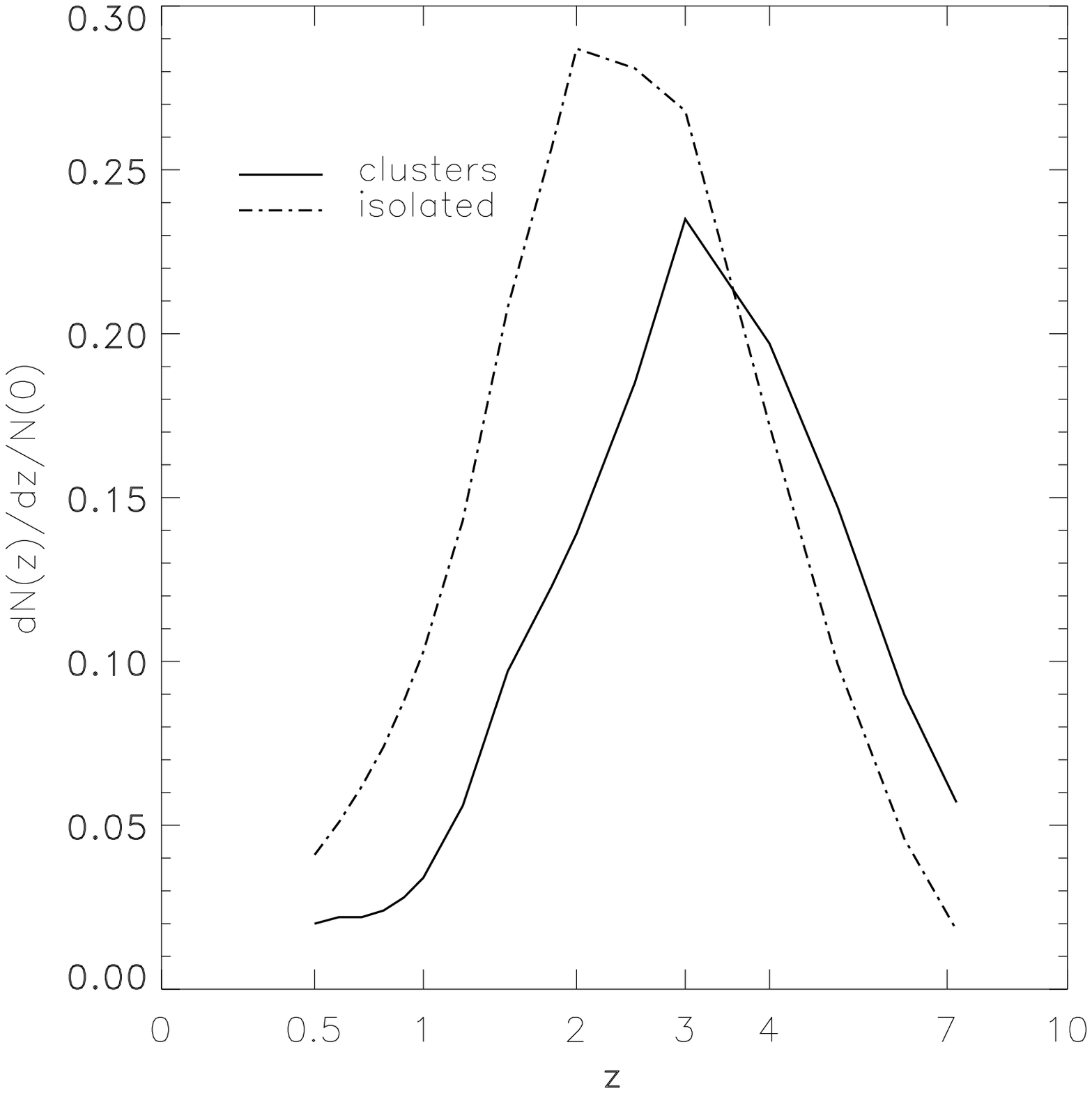}}

\rput[tl]{0}(0.,3.5){
\begin{minipage}{8.5cm}
  \small\parindent=3.5mm {\sc Fig.}~3.--- The distribution of
  ``formation redshifts'' defined as a redshift when the most massive
  progenitor of the corresponding present-day halo reaches the maximum
  circular velocity higher than 50 km/s. All the $z=0$ halos with the
  maximum circular velocities in excess of 120 km/s were selected.
  This distribution can be interpreted as a distribution of redshifts
  of $z=0$ halos with circular velocities $> 120$ km/s at which they
  become capable of hosting a luminous galaxy.
\end{minipage}
}
\endpspicture}
halos as halos consisting  of 3 or more subhalos  and having masses of
$\lesssim 10^{14} \Mhsun$.  Two halos located  within a common cluster
of overdensity 200,  are considered to be {\em  pairs}.  In the  subsequent
analysis, we  have omitted all  pairs due to  the following reasons. A
massive  halo with  a single  subhalo   of much  smaller  mass  should
probably be considered  isolated.  Subhalos  of even
smaller mass could  have been unresolved or missed  due to the limited
mass resolution so that such pair, depending on mass, should have been
a small group  rather than an isolated  halo.  To avoid this  kinds of
confusing  identifications  of the  environment, we  have  omitted all
pairs. The analyzed halos,    therefore, are classified   as isolated,
cluster, or group halos.

The procedure described above results in identification at $z = 0$ in
our $60 h^{-1}$ Mpc box of 401 cluster halos (10 \%) in 18 clusters,
743 halos in groups (18 \%) and 2545 isolated halos (60 \%).  The
remaining 504 halos are found in pairs (12 \%).  The first cluster has
formed between $z=2.5$ and $z=2$. The fraction of galaxies in clusters
increases with time, whereas the fraction of isolated galaxies in the
considered mass range decreases, and the fraction of pairs remains
approximately constant (Gottl\"ober \etal 1999c).

\subsection{Progenitors of halos}
\label{def_prog}
For each of the halos in our $z=0$ sample we have constructed a
complete evolution tree over 25 epochs ($z = 0$, 0.05, 0.1, 0.15, 0.2,
0.25, 0.3, 0.4, 0.5, 0.6, 0.7, 0.8, 0.9, 1.0, 1.2, 1.5, 1.8, 2.0, 2.5,
3.0, 4.0, 5.0, 6.1, 7.2, 10.0). Procedure of progenitor identification
is based on the comparison of lists of particles belonging to the halos
at different 
{\pspicture(0,0)(11.5,12.5)

\rput[tl]{0}(0,12.5){\epsfxsize=8.5cm
\epsffile{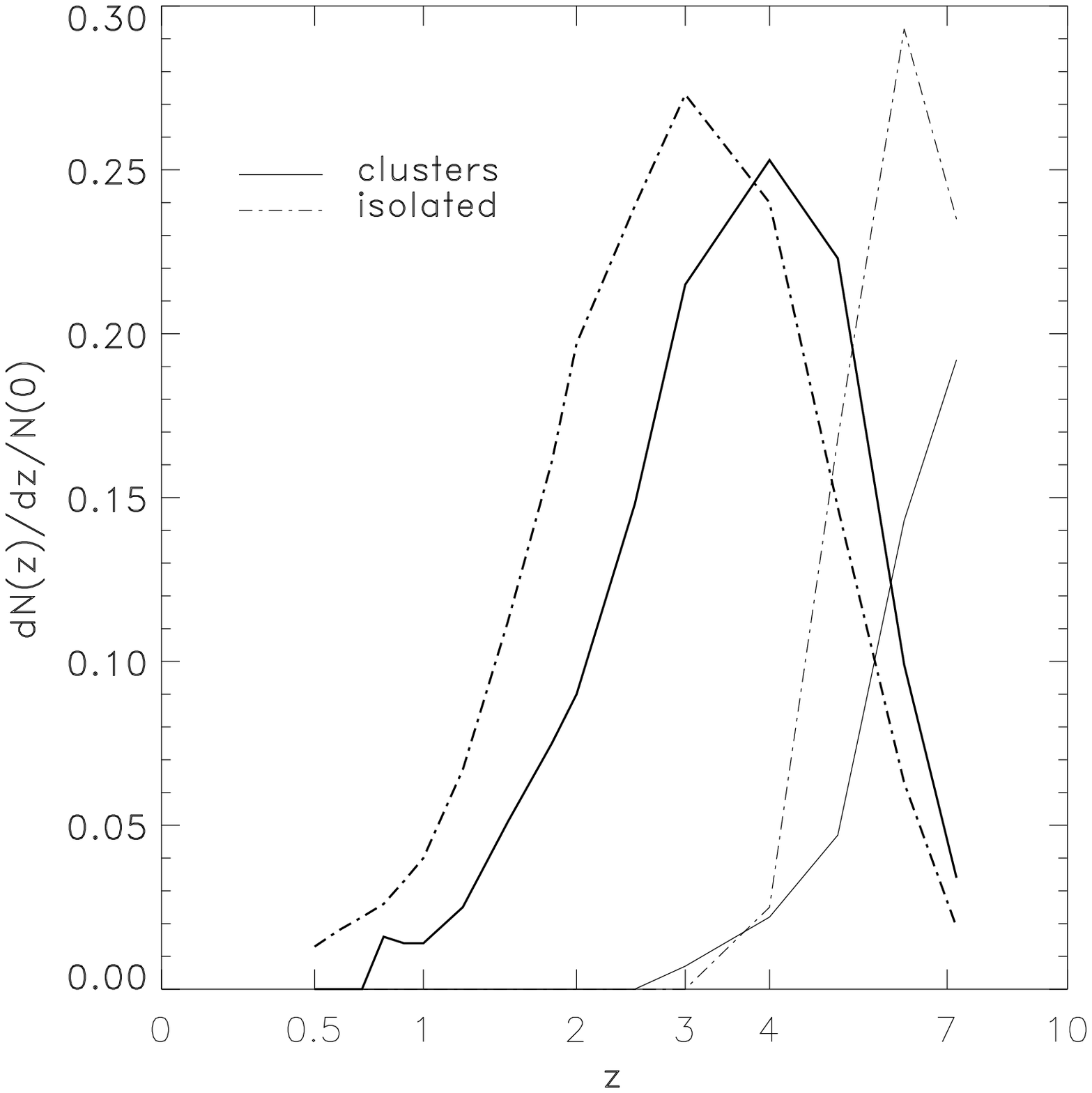}}

\rput[tl]{0}(0.,3.5){
\begin{minipage}{8.5cm}
  \small\parindent=3.5mm {\sc Fig.}~4.--- The same as
  Fig.~3 but for more massive halos. {\em Thick
    lines}: A subsample of $z=0$ halos with $150 > v_{circ} >
  200$~km/s.  {\em Thin lines}: a subsample of $z=0$ halos with
  $v_{circ} > 300$~km/s.
\end{minipage}
}
\endpspicture}
redshifts both back and forward in time (for details see
the Appendix). Our algorithm of tracing halo histories identifies the
correct ``ancestor-descendant'' relationships rather accurately, with
estimated ancestor-descendant misidentifications ${_ <\atop{^\sim}}
2\%$ of the cases (Gottl\"ober \etal 1999a). These misidentifications
happen usually with small-mass halos consisting of only a few tens of
particles (below the mass limit of halos in our sample), \ie the halos
in the mass range of our sample are not significantly affected.

Now we define the {\em halo detection epoch} as the epoch at which the
halo has reached a threshold when it can be expected to host a galaxy.
The threshold is taken to be $v_{circ} > 50$  km/s for the first time. 
Our halo-finder  assumes a minimum  circular velocity, $v_{circ} > 50$
km/s, and  a  minimum  number  of   bound particles  of  40,  for  the
progenitors of our halos at  $z>0$. In Fig.~3 we  show
how many cluster and     isolated halos were detected   per   redshift
interval as a function   of redshift.  In  this  figure we  show  only
cluster  halos; halos  in groups  show similar  behavior.  Due  to the
construction   procedure   of our    sample   (\S~3.1), all
present-day  halos exist already  at $z = 0.3$. Fig.~3
shows that in general cluster halos  form earlier than isolated halos. 
The maximum formation  rate  is reached at $z  \approx  3$ whereas the
maximum of  the formation rate   of isolated  halos is  reached  later
($z<2.5$).   Cluster halos form in   regions of higher overdensity and
therefore reach the detection threshold earlier.

The integral over the  curves  of Fig.~3. gives us  the
completeness of the  progenitor samples as  a function of redshift. It
reaches  90 \% (50 \%)  at $z=2$ ($z=4$) for halos  in 
{\pspicture(0,0)(11.5,12.5)

\rput[tl]{0}(0,12.5){\epsfxsize=8.5cm
\epsffile{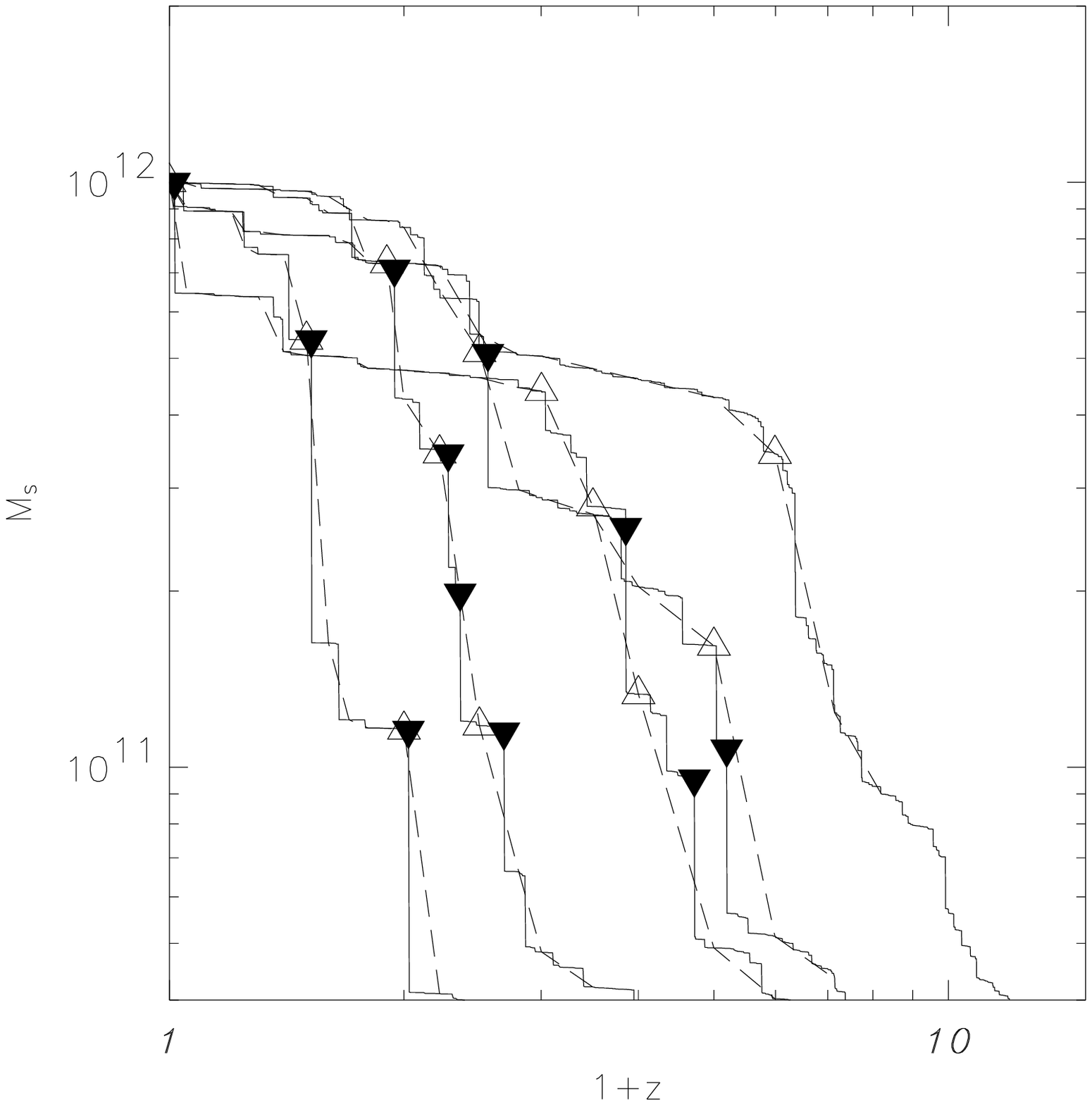}}

\rput[tl]{0}(0.,3.5){
\begin{minipage}{8.5cm}
  \small\parindent=3.5mm {\sc Fig.}~5.--- Test of effects of finite
  time interval used to identify major mergers.  {\em Solid triangles}
  denote major merger events found in five EPS merger histories with
  very fine time resolution ({\em solid lines}).  The {\em dashed
    lines} show the mass evolution of the corresponding objects
  sampled using time intervals similar to those used in the analysis
  of the simulation. {\em Open triangles} denote major merger events
  detected using these redshift intervals. At low redshifts ($z\la
  2$), the effect of finite sampling is negligible.
\end{minipage}
}
\endpspicture}
clusters and at
$z=1.5$ ($z=3$) for isolated halos.

Let us consider a subsample of halos with  $150 > v_{circ} > 200$~km/s
(Fig.~4, thick lines). Cluster  halos of this  sample
form  earlier  ($z \sim   4$) than isolated   halos.  We see the  same
tendency for the subsample of the most  massive halos with $v_{circ} >
300$~km/s (thin lines), however  due to poor statistics  we do not see
the  maximum of formation  rate  of these  massive cluster halos.  The
progenitor of the  central  halo of  the  most massive  cluster (which
corresponds  to a  massive central  cD galaxy)   can be  identified at
$z=15$ as an object of $3\times 10^{10} \Mhsun$.

\subsection{Extended Press-Schechter formalism}\label{eps}

To test how sensitive the merger rate estimates are to our
assumptions, we use the synthetic merger histories generated
using the extended Press-Schechter formalism (hereafter EPS, Bond et
al. 1991; Bower 1991; Lacey \& Cole 1993). Specifically, we use the
``$N$-branch trees with accretion'' method of Somerville \& Kolatt
(1999) to construct merger histories for the host. 

In this method the progenitors of a halo of mass $M_h$ at a given epoch
and in mass range $[M_p^{min},M_h]$ are determined through a series of
Monte-Carlo picks using probability for a halo to have accreted mass
$\Delta M$ during time $\Delta t$ (see eq. 2.29; of Lacey \& Cole
1993).  $M_p^{min}$ is the minimum progenitor mass that is kept track
of, specific to particular intended use of the merger histories.  The
method was slightly modified; at each step we require the number of
progenitors in the mass range of interest to be close to the expected
average.  This modification significantly improves
{\pspicture(0,0)(14.5,15.5)

\rput[tl]{0}(0,15.5){\epsfxsize=9.cm
\epsffile{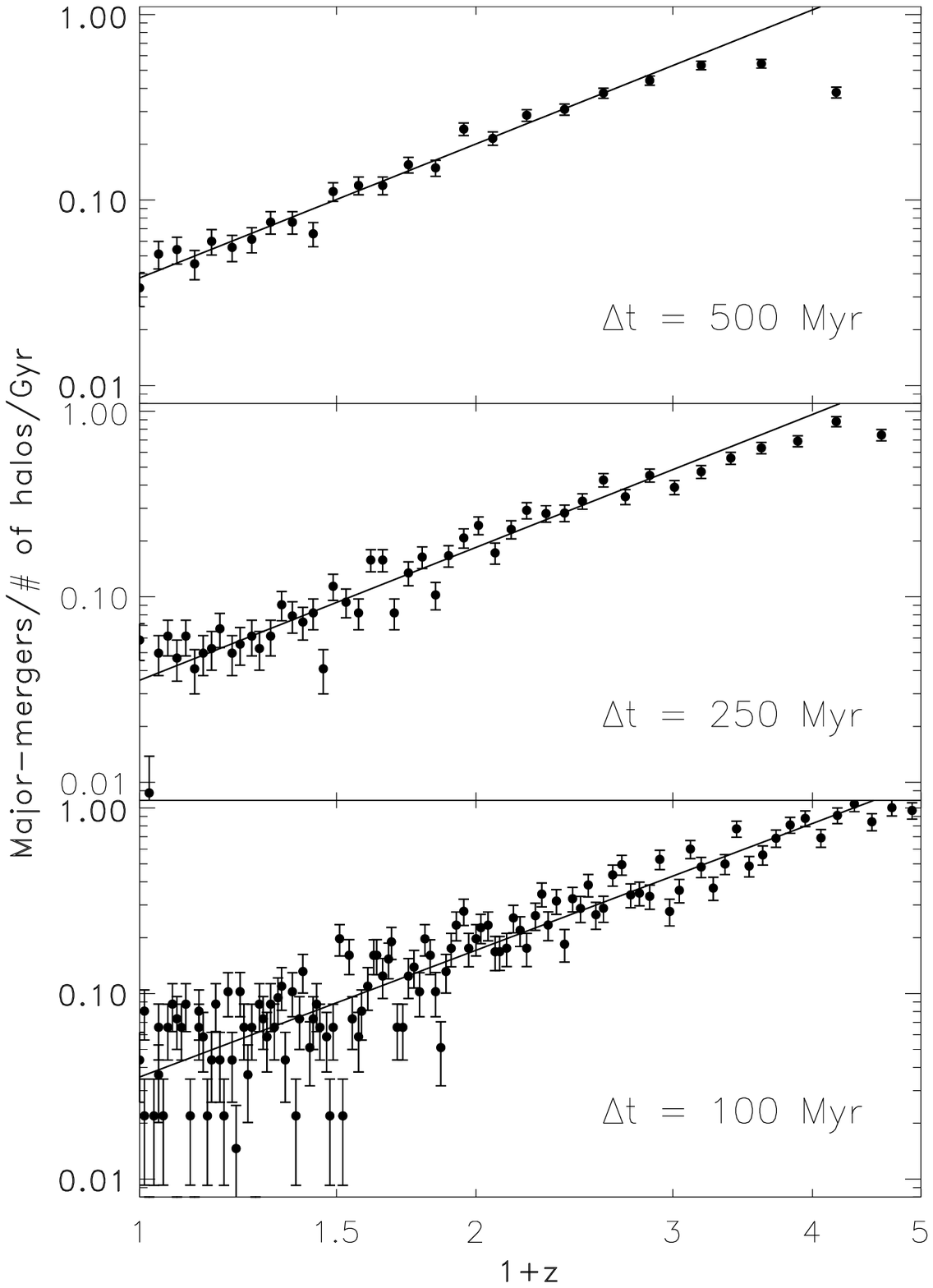}}

\rput[tl]{0}(0.,3.5){
\begin{minipage}{8.5cm}
  \small\parindent=3.5mm {\sc Fig.}~6.--- Merging rates estimated
  using the EPS merger histories using the same procedure that we used
  to estimate the merging rate in the simulation. The $y$-axis show
  the number of major mergers per Gyr at a given redshift normalized
  by the number of progenitors at this redshift. The {\em solid lines}
  are power law fits ($\propto (1+z)^{\alpha}$) to the $z < 2$ points.
  The mergers are detected in time intervals of 500 Myr (top panel,
  $\alpha = 2.40$), 250 Myr (middle panel, $\alpha = 2.38$), and 100
  Myr (bottom panel, $\alpha = 2.27$) time intervals.
\end{minipage}
}
\endpspicture}
agreement of the
progenitor mass function generated by the method with the analytical
prediction (Kravtsov et al. 2000).  We refer reader to Somerville \&
Kolatt (1999) for further details of the method.

For each step in time, a merger history contains information about the
mass of the host at the current epoch and masses of its progenitors at
the previous epoch.  The most massive progenitor in the list is assumed
to represent the host halo a time step back in time, while the other
progenitors are considered to be the halos accreted during the
step. The time steps were chosen as prescribed by Somerville \& Kolatt
(1999).

Using the EPS method we have generated a sample of 1350 merging
histories of halos of mass $10^{12} \Msun$ at $z=0$. This is a typical
mass for halos in our numerical catalogs. We followed merger histories
back in time until the mass of the most massive progenitor falls below
$4 \times 10^{10}\Msun$, the same minimum mass that was used in
analysis of the simulation. With a mean $\Delta z $ of about 0.003 the
shortest history (425 steps) starts at $z=1.4$, whereas the longest
(3802) starts at $z=11$. Our goal is to use the high temporal
resolution of the EPS merging histories to test the effect of the
coarse temporal resolution of the merger histories constructed using
simulation. 

As discussed in \S~2, we define a major merger as an event
in which the mass of a halo increases by more than a certain threshold
in a given time interval.  In Fig.~5 we show five of the
original EPS merger histories (solid lines) and the major merger events
detected (solid triangles) using the original time steps of the history
and a high threshold of 0.35. The dashed lines show the evolution of
the same halos tracked only at 25 time moments used in the simulation
(see \S~3.3) where the open triangles denote the major
merger events detected for this history using the same definition of
the major merger as before. One can see that at $z\la 2$ each solid
triangle is accompanied by an open one.  At higher $z$ there are
exceptions. In one of the histories two subsequent merger events
detected in the original merger history are detected as only one event
in the coarse-time history. In another history (rightmost curve) two
successive minor mergers in the original history (at $z\approx 5$),
that are not classified as ``major'' add up to a major merger when the
time resolution is degraded.  The fact that two (or more) successive
minor mergers occuring within less than 0.5 Gyrs are treated as one
major merger is not necessarily wrong.  The physical effect of large
mass accretion within a short time interval may be the same regardless
of whether this accretion was through a single major or several minor
mergers. However, the fact that coarse temporal resolution may
masquerade several major mergers as one, may lead to an underestimate
of the merger rate at $z\ga 2$.

In the following we want to test  whether the estimates of the merging
rate are influenced  by the relatively coarse  time intervals  used in
simulation  analysis. To this end, we  tracked the original EPS merger
histories  in equally spaced time intervals  of 500  Myr, 250 Myr, and
100  Myr.  We have then  counted the major  mergers as events in which
mass increases by  more than 25\% in  a given time interval.  Finally,
we have calculated  the merging rate of the  EPS halos in the same way
as for the halos of the simulation:  the number of major merger events
per halo per Gigayear. The result is shown in the three panels of Fig.~6.
  The  solid  line  is  a  power    law fit  ($\propto
[1+z]^{\alpha}$) for $z < 2$ epochs,  with $\alpha = 2.40$ for $\Delta
t = 500$ Myr, $\alpha = 2.38$ for $\Delta t  = 250$ Myr, and $\alpha =
2.27$ for $\Delta t = 100$ Myr.

There is a good agreement of the merging rates for different time
intervals. The somewhat lower merging rate in case of $\Delta t = 100$
Myr is a result of the poor statistics at low $z$. Obviously, the
merging rate increases with redshift, but it cannot increase to
infinity due to the adopted major merger definition.  Clearly, if one
assumes that all halos underwent a major merger in the given time
interval $\Delta t$ (the algorithm by design cannot detect multiple
major merger events in a particular time interval, see above), the
merging rate reaches its maximum value $1/\Delta t$, i.e. 2, 5, or 10
in the three panels of Fig.~6. One can see that the
curves in this figure indeed flatten before reaching this value, but at
different epochs.  This flattening is due to the limited mass
resolution of the simulation. At a given time moment and with a given
resolution, some of the halos fall below the detection threshold at the
earlier time moment, i.e. for these halos major merging events cannot
be detected.  With larger $\Delta t$ this probability increases and
reduces therefore the rate and epoch at which this effect sets in.
However, Fig.~6 shows that for the 
{\pspicture(0,0)(14.5,17.5)

\rput[tl]{0}(0,17.5){\epsfxsize=9cm
\epsffile{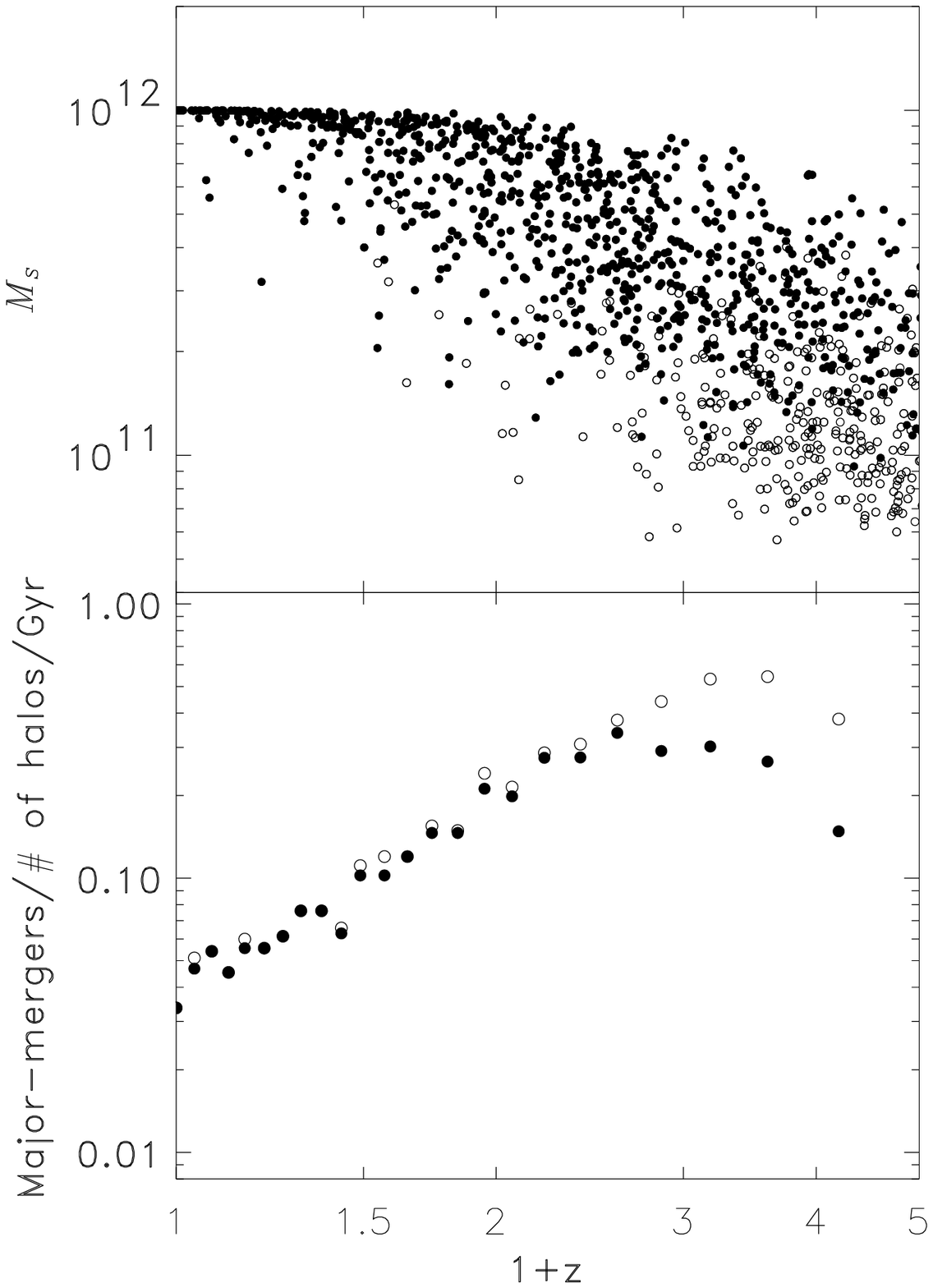}}

\rput[tl]{0}(0.,5.){
\begin{minipage}{8.5cm}
  \small\parindent=3.5mm {\sc Fig.}~7.--- Comparison of major merger detections using different definition 
  of the major merger events. {\em Top panel:} the circles (both solid
  and open) denote all major mergers detected for 1000 EPS merger
  histories in time intervals of 500 Myr (mass growth threshold of
  0.25). The {\em solid circles} denote the major mergers of two
  massive progenitors (halos of comparable mass), while the {\em open
    circles} denote the accretion of multiple small halos. {\em Bottom
    panel:} the merging rate evolution corresponding to the two
  different merger definitions. Note that the two definitions result
  in a similar merger rate estimates at $z\la 2$. 
\end{minipage}
}
\endpspicture}
interval of $\Delta
t=0.5$ Gyr (the interval used in our simulation analysis), the results
are robust for $z\lesssim 2$.

In Fig.~7 we compare two different detection schemes
of major merger events. Open   circles denote all major merger  events
detected  by  the condition  $(M_2 -  M_1)/M_2 >   0.25$. These events
include  events due to  the merger  of two  massive halos and multiple
mergers  and/or rapid   accretion.    The filled  circles  denote  the
subsample of binary mergers of  two massive progenitors.  Note that in
the  upper panel almost all open  circles coincide with filled circles
at  $M  > 2  \times 10^{12}   \Mhsun$,  i.e. the  major  merger events
detected through the mass  growth are in fact due  to binary mergers.  
With  decreasing  mass  and increasing  redshift  the  number of major
mergers with  only one massive progenitor   rapidly increases. This is
simple  due to  the  fact that  with decreasing mass  of  the halo the
probability that its second   massive progenitor is already below  the
detection\\ 
{\pspicture(0,0)(7.,8.)

\rput[tl]{0}(0,8.){\epsfxsize=8.5cm
\epsffile{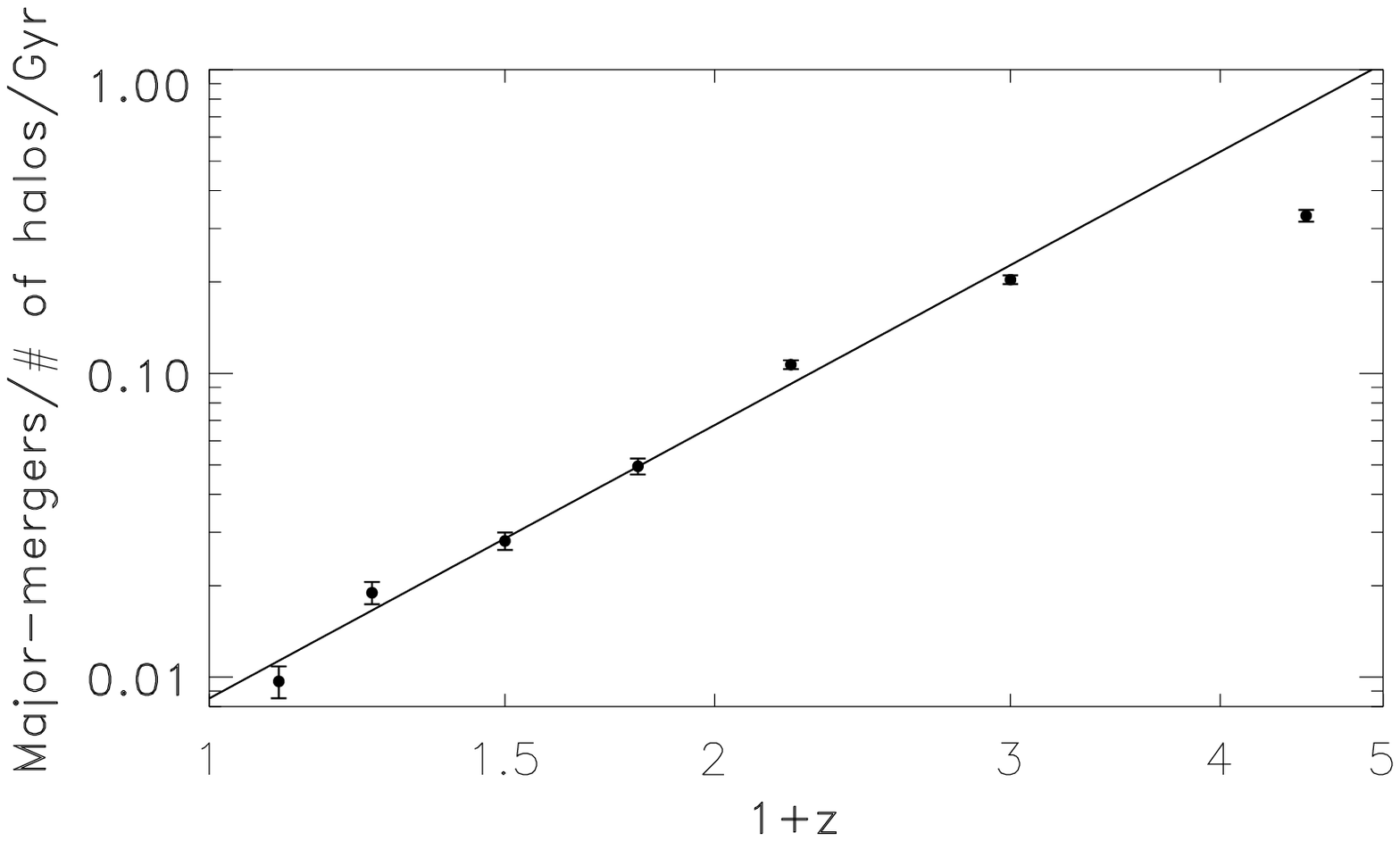}}

\rput[tl]{0}(0.,2.5){
\begin{minipage}{8.5cm}
  \small\parindent=3.5mm {\sc Fig.}~8.--- Number of major mergers
  identified in the simulation at a given redshift and normalized to
  the number of all most massive progenitors at this redshift. The
  {\em solid line} shows a power law fit ($\propto (1+z)^{3.0}$) to
  the first six points ($z<2$).
\end{minipage}
}
\endpspicture}\\
threshold increases. Although in the EPS formalism one could
easily extend the tree to lower masses,  in simulations one is limited
by    the   mass   resolution.    In  the     lower panel  of     
Fig.~7 we plot the merging  rates detected for all  major
mergers  (open  circles,    the same as     the top  panel of   Fig.~6) 
and  for binary major  mergers (filled circles). There
is no  differences at  $z\lesssim  2$.  We conclude   that there is no
difference between the two  definitions at redshifts $z\lesssim 2$. It
is  more reasonable to analyze  the simulation using  the mass growth,
$(M_2  - M_1)/M_2$,  as   a  major  merger  definition,  because  this
procedure provides informations to somewhat larger redshifts (see 
Fig.~7,  lower panel).  More investigations are necessary
to understand whether the bend at higher $z$ is real.

\section{RESULTS AND DISCUSSION}\label{discuss}

As was   noted above,  to identify major    mergers  we calculate  the
relative   mass growth $(M_2  -  M_1)/M_2$   during the time  interval
$t_2-t_1$.  Due to the  selection  of simulation output redshift,  the
time  intervals  (see   Sect. \ref{def_prog})  have  somewhat variable
length. As a reminder, we assume a major merger to have occured if the
relative mass   growth  is larger  than  0.25.   As  discussed in  the
previous section,  we calculate  the  {\it total}  change of mass, not
only the  contribution of  merging  with another massive  halo. In the
previous  section we  have shown that  variable  time intervals do not
change the result if this {\it merger  rate} is normalized
by time interval. Therefore, we  calculate  the {\it  merger
  rate} as  the number of major-merging events  of  the progenitors of
our halo sample per gigayear  normalized to the number of  progenitors
at a given moment. In Fig.~8  we show the evolution of
the  merger  rate in this   definition.  The estimated  merger rate is
obtained by averaging over three subsequent time intervals in order to
reduce the  scatter.  The error bars   are $\sqrt{N}$ errors   for the
number of events detected.

For redshifts  $z \la 2$  our  sample is  90  \%  complete. For  these
redshifts  the merger   rate  can be  fitted   by a simple  power  law
$(1+z)^{3.0}$. There is   an indication of  flattening of  the merging
rate at higher  redshifts. As shown  in Fig.~3 at $z >
2$ we are rapidly loosing the halo progenitors due to mass resolution.
Moreover, Fig.~6 also shows that this effect is likely to
be due to the limited mass resolution. However, as we have argued above, 
prediction for the evolution at $z\la 2$ is reliable. 

Recently,   Le F\`{e}vre \etal  (1999)   published  the first   direct
observational measurement of the merger fraction at redshifts $z>0.5$.
They have used visual merger identifications as  well as statistics of
close  pairs  of galaxies.  To transform the  observed merger fraction
into a merger rate requires knowledge of the lifetime of a merger, \ie
the time for which the traces of the merging event can be observed. An
upper limit of 0.4 to 1  Gyr, as assumed by  Le F\`{e}vre \etal (1999),
approximately corresponds  to  the time  step which   we  have used to
derive the   merging rate. Le  F\`{e}vre \etal  (1999)  have derived a
merger rate varying   with  redshift as $\propto  (1+z)^{3.2   \pm
  0.6}$. This result    is  in good  agreement with    our theoretical
prediction.

In principle,  the evolution  of the merger  rate  is a very important
observable for testing  cosmological   models. In practice,  both  the
theoretical and  observational estimates of merging  rates depend on a
number of  assumptions.  In our model  estimates,  the  result depends
mainly on the   definition (mass threshold)   of major merger events.  
Increasing  the threshold  0.25 to  0.4,    corresponds to  a  faster
evolution  described   by  $\propto  (1+z)^{3.7}$.  Lowering  the  
threshold to 0.20, results in a slightly slower evolution of the merger 
rate. Observationally, merger rate evolution is determined by studies of
the evolution of the correlation function, the evolution of pair
numbers or by morphological studies (e.g., Abraham
1999). In the first two cases the merging rate is not directly measured
and the conversion of the measured quantities to the merging rate is
not straightforward. The third approach, \ie observing mergers in
progress, is the most direct one (cf. Le F\`{e}vre \etal 1999). 

From an excess of power in  the observed two-point angular correlation
function at     angular  scales  $2^{\prime\prime}    \le  \theta  \le
6^{\prime\prime}$ Infante \etal (1996)   determined a merging rate  of
galaxies $(1+z)^{2.2 \pm  0.5}$. According to  Burkey \etal (1994) the
pair fraction  in deep HST images  grows with  redshift as $(1+z)^{3.5
  \pm  0.5}$  from  which they deduce  a  merger  rate $(1+z)^{2.5 \pm
  0.5}$.  On the  contrary, Carlberg \etal   (1994) conclude from  the
evolution of the  close pair ($\theta \le  6^{\prime\prime}$) fraction
that  the merger rate--redshift  relation  is $(1+z)^{3.4  \pm  1.0}$. 
Finally, Yee  \& Ellingson (1995) found  for the close pair (projected
distance   less than $20  \hkpc$)   evolution rate  a somewhat steeper
redshift  dependence $(1+z)^{4.0 \pm  1.5}$.   Different authors  used
different relationships between the evolution of pair fraction and the
evolution of the merger  rate. Nevertheless, the derived merger rates 
are  in general  agreement and  also   in agreement with  our
theoretical predictions.

It is very interesting to see whether the merger rate evolution
depends on the halo environment. In Fig.~9 we show the
merger rate of cluster, group and isolated halos normalized to the
merger rate of all halos shown in Fig.~8. The higher
rate of major mergers at early epochs for cluster and group halos is
due to the higher density in regions, where cluster and groups have
been forming.  Note that the clusters have not yet existed at
$z\ga 2-3$. As clusters with large internal velocities form,
merging rate quickly decreases (Makino and Hut 1997; Kravtsov \&
Klypin 1999; Mamon 2000).  There are almost no major merger events of
cluster halos in the recent past. Those 8 events at $z = 0.1,\; 0.25$,
and 0.5 have probably happened just outside
{\pspicture(0,0)(11.5,12.5)

\rput[tl]{0}(0,12.5){\epsfxsize=9cm
\epsffile{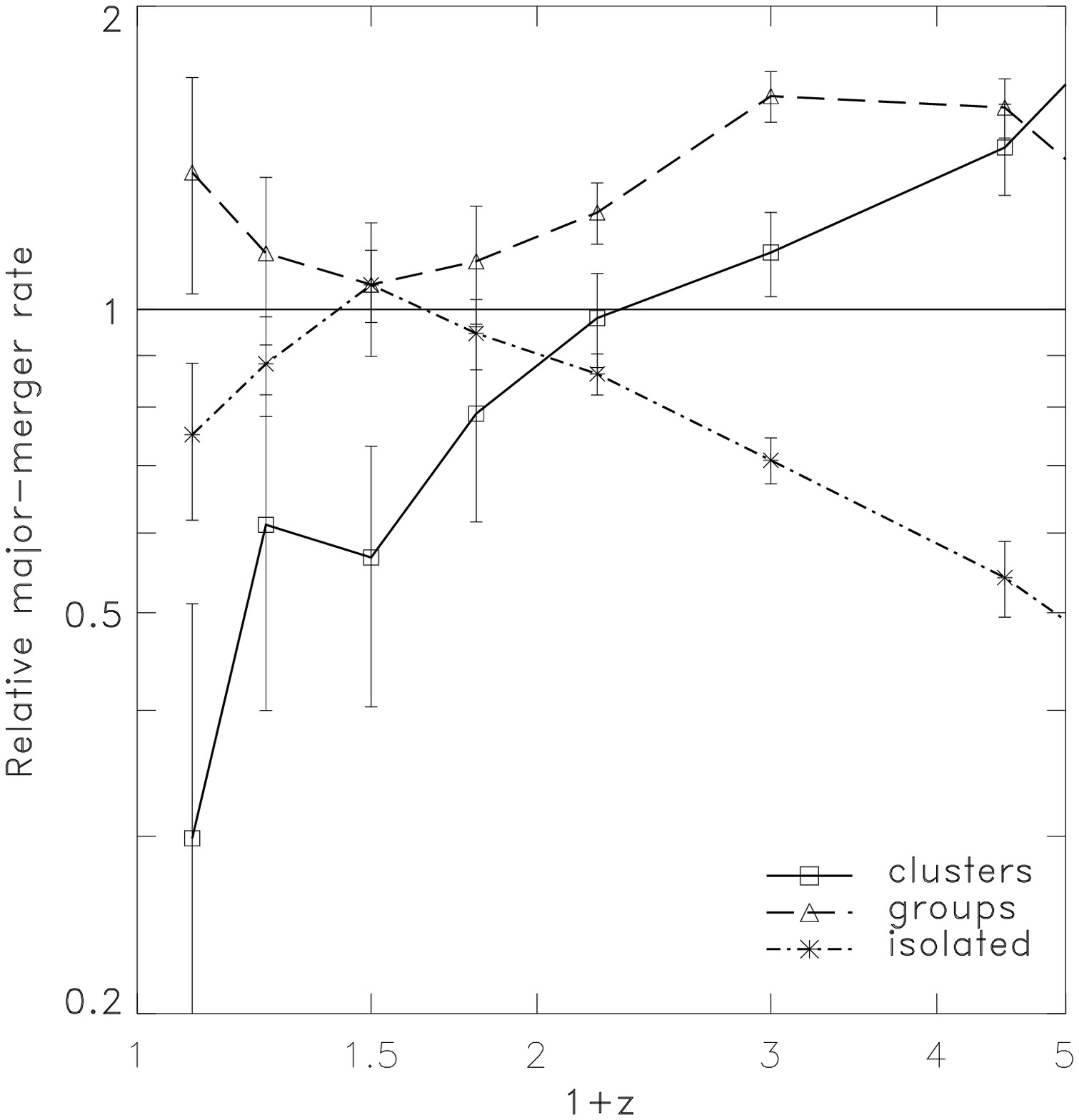}}

\rput[tl]{0}(0.,3.5){
\begin{minipage}{8.5cm}
  \small\parindent=3.5mm {\sc Fig.}~9.--- The relative merging rate of
  the population of cluster, group, and isolated halos. We plot the
  ratio of the merging rate of particular population to the merging
  rate of all major progenitors at given redshift. Values larger than
  unity imply that this population has a higher rate than the overall
  population.
\end{minipage}
}
\endpspicture}
the clusters before the
halos were accreted by cluster or, alternatively, they might have
occured within the surviving low internal velocity substructute within
cluster. Note also that the massive central halo of the cluster
(which should correspond to the observed brightest cluster galaxies) may
accrete other halos after the cluster has been formed (Dubinski 1998;
Mamon 1999). If the accreted halo is massive enough a major merging
event would be detected. The low merger rate in simulated clusters has
been noted also by Ghigna \etal (1998), although mergers within accreted 
substructure have been observed (Springel et al. 2000).

\section{CONCLUSIONS} \label{concl}

We have analyzed a high-resolution collisionless simulation of the
evolution of structure in a $\Lambda$CDM model. We have followed the
formation and evolution of DM halos in different cosmological
environments and estimated the evolution of the major merger rate of
dark matter halos.

We have found that regardless of their present-day mass, halos that
end in clusters form earlier than isolated halos of the same mass
(Figs.~3 and 4). We find that at
redshifts $z \la 2$ major merger rate evolves as $(1+z)^{\sim 3.0}$, in
good agreement with observations.  Finally, we have calculated the
merging rate evolution as a function of halo environment at $z=0$
(Fig.~9).  The merger rate of halos located in
clusters or groups at present increases faster back in time than that
of isolated halos. The cluster and group halos are therefore predicted
to have a higher rate of major merger events in the past. At $z\la 1$,
the merger rate of cluster and group halos drops very quickly, while
numerous major merger events for isolated halos have been detected
down to $z=0$. This implies possible systematic differences between
cluster and field ellipticals. Evidence for such differences was found
by de Carvalho \& Djorgovski (1992), while Bernardi \etal 1998
detected close similarity between cluster and field early type
galaxies.

The agreement between theoretical predictions and observations are
encouraging and supports the validity of the hierarchical structure
formation scenario. Future, higher resolution simulations should
extend the predictions presented here to higher redshift and, in case
of gasdynamics simulations, provide a more straightforward connection
to observations. On the observational side, the ever increasing size
of the high-redshift galaxy samples should also allow estimates of the
merger rate at high redshifts in the near future.

\acknowledgments This work was funded by the NSF and NASA grants to
NMSU.  SG acknowledges support from Deutsche Akademie der
Naturforscher Leopoldina with means of the Bundesministerium f\"ur
Bildung und Forschung grant LPD 1996.  A.V.K. was supported by NASA
through Hubble Fellowship grant HF-01121.01-99A from the Space
Telescope Science Institute, which is operated by the Association of
Universities for Research in Astronomy, Inc., under NASA contract
NAS5-26555.  We acknowledge support by NATO grant CRG 972148.

\appendix{\bf Appendix:  Technical details}

{\bf A: Halo identification:} Our halo identification algorithm (see
Klypin \etal 1999 for more details) starts with the search of local
density maxima assuming a smoothing radius of the order of $\sim 10
\hkpc$. This radius defines the scale of the smallest objects we are
looking for.  Once the centers of potential halos are found, we start
the procedure of removing unbound particles and finding the size of
halos. We determine the mass of the dark matter particles in
concentric spherical shells around the halo center, the mean shell
velocity, and the velocity dispersion relative to the mean. We assume
that particles with velocities larger than the escape velocity at the
position of the particle are not bound to the halo and do not take
them into account when calculating halo properties.  Using the
density profile of the halo we estimate the maximum rotational
velocity and the radius at which the maximum is reached. We will call this 
radius, radius of the halo. 

Removal of unbound particles is important in the case when a halo with
a small internal velocity dispersion moves inside a larger halo. In many 
cases, the small-mass subhalos can be unambigously identified only
after removing the unbound background particles of the larger halo. 
To avoid misidentifications of group or cluster halos as galaxy halos we have
introduced a maximum possible halo radius of $100 \hkpc$. We assume 
this radius to be the halo radius if the circular velocity profile is still 
increasing at this distance. 
   
The halo identification scheme is crucial for the algorithm which
finds the most massive progenitor of a halo. Due to our halo
definition, DM particles can belong to more than one halo (e.g., a particle
may be gravitationally bound to both the subhalo and to its host halo)
At a given time moment, some of the particles 
gravitationally bound to a halo might be outside of the formally defined halo
radius.  Both situations must be taken into account if one constructs the
evolution history of halos using lists of particles
belonging to a given halo.

We use the constructed density profile and estimated radius to assign
each halo a mass and circular velocity.  To construct halo catalogs,
we select all halos with maximum circular velocities above a certain
minimum value.  In addition, we have rejected all halos with the
number of bound particles below a minimum threshold.

There are also some special cases in halo identification. For example,
after a recent merger two local density maxima could be found within a
common halo. The halo finder tends to identify such a configuration as
two halos at a distance of the order or smaller than the radius of the
halos.  Also, at a given moment small clumps in a dense environment
could, by chance, appear as bound clumps which, however, disappear by
the next time moment. These misidentifications could masquerade as
recent merging events. To avoid these kinds of misidentifications, we
have required that each halo of the sample at $z=0$ has a unique
progenitor at the last five time steps (see \S~3.3). If a
progenitor has been already identified as progenitor of another halo,
the smaller mass halo is discarded.  This procedure, which removes
about 6\% of the halos, reduces the number of fake merger detections
and makes the results more reliable.

{\bf B: Definition of environment:} We define the environment of a
halo using the mass of the virialized host (if any) to which the halo
belongs. We find virialized systems using a friends-of-friends
algorithm with a linking length of 0.2 times the mean density. For
each halo, we determine the particle with the shortest distance to the
center of mass of the halo. Then we search for the virialized object to
which the particle belongs. Since these objects have overdensities of
the order 200, we are confident that our halo of higher internal overdensity
belongs to a virialized host if the most central particle belongs to it.

{\bf C: Progenitor identification:} We begin by identification of all
particles bound to a halo at a given redshift $z_i$. Then we find all
objects (identified halos, small groups of particles below the
detection limit of halos, isolated particles) at the previous redshift
$z_{i+1}$ which contain any of the particles of the halo in question. The
time interval between the two redshifts is typically of the order of
0.5 Gyr. We repeat this procedure for all halos at a given
redshift. We thus obtain complete information on the origin
of the particles found in
halos at redshift $z_i$. However, due to the halo
identification procedure described above this information is {\it not}
equivalent to the mass of the progenitor objects. Indeed, we have
determined the mass at the radius of the maximum rotational
velocity. To get the mass growth due to merging and accretion we must
compare the masses of the halo and its progenitor which are defined in
the same way. 

It is relatively straightforward to identify the most massive
progenitor for isolated halos: it is simply the halo which contains
the largest fraction of particles of the descendant. In a dense
environment, on the other hand, the identification is more
complicated. For example, the particles of a subhalo belong both to
the satellite halo and the hosting halo at an earlier moment, so that
both would be identified as progenitors. To avoid this (and similar)
misidentifications we identify not only the ancestors of all halos
found at $z_i$ but also the descendants of all halos found at
$z_{i+1}$ and check whether a given halo is really the descendant of
its ancestor by searching for the maximum subsamples of particles
belonging to the corresponding halos. With such somewhat extensive
procedure we reduce the misidentifications of progenitors to ${_
  <\atop{^\sim}} 2\% $ of all considered cases. Note, that these
misidentifications lead to scatter in the mass evolution history.

\end{document}